\documentclass[11pt,leqno]{article}

\usepackage{graphicx}
\usepackage{amsmath, amsthm}
\usepackage{bm}
\usepackage{rotating}
\numberwithin{equation}{section}

\usepackage{hyperref}

\begin{document}

\newcommand{\bea}{\begin{equation}}
\newcommand{\eea}{\end{equation}}
\newcommand{\longeq}{\;\;=\;\;}
\newcommand{\fatb}{\mathbf{b}}
\newcommand{\fate}{\mathbf{e}}
\newcommand{\fatf}{\mathbf{f}}
\newcommand{\fatg}{\mathbf{g}}
\newcommand{\fati}{\mathbf{i}}
\newcommand{\fatk}{\mathbf{k}}
\newcommand{\fatn}{\mathbf{n}}
\newcommand{\fatq}{\mathbf{q}}
\newcommand{\fatr}{\mathbf{r}}
\newcommand{\fatu}{\mathbf{u}}
\newcommand{\fatv}{\mathbf{v}}
\newcommand{\fatw}{\mathbf{w}}
\newcommand{\fatx}{\mathbf{x}}
\newcommand{\faty}{\mathbf{y}}
\newcommand{\fatz}{\mathbf{z}}
\newcommand{\fatS}{\mathbf{S}}

\title{Derivation of the fundamental equations of continuum thermodynamics from
statistical mechanics\footnote{This study emerged from a report on a
work of {\sc Irving} and \textsc{Kirkwood} \cite{irki:50} that was
presented by the author in a seminar at Indiana University in summer
1954. The seminar was organized by Professor {\sc Clifford
Truesdell} whose inspiration is gratefully acknowledged by the
author.}\\\textsl{Translation}\footnote{by Rich Lehoucq abd O. Anatole von Lilienfeld, Sandia National Laboratories. This translation is for research purposes and must always properly cite the original article. Corrections to the original paper listed in an appendix. Sandia is a multiprogram laboratory
operated by Sandia Corporation, a Lockheed Martin Company, for the
United States Department of Energy under contract
DE-AC04-94AL85000.} of the paper\\
Die Herleitung der Grundgleichungen der Thermomechanik der
Kontinua aus der Statistischen Mechanik, Indiana Univ. Math. J.,
4:627--646, 1955.\footnote{Article originally published in the \emph{J. Rational Mech. Anal.}. The article is available at digital archives of the Indiana University Mathematics Journal (IUMJ) located at \href{http://www.iumj.indiana.edu/IUMJ/FULLTEXT/1955/4/54022}{http://www.iumj.indiana.edu/IUMJ/FULLTEXT/1955/4/54022}.}}
\author{WALTER NOLL}
\date{}

\maketitle

\noindent \textbf{Introduction.} Assuming a classical statistical
system of point particles the fundamental equations of continuum
thermomechanics (continuity equation, equation of motion, and energy
equation) shall be derived exactly. The macroscopic state functions
(density, velocity, stress, energy density, heat flux) are
interpreted as expected values.

The systems can represent any continua (gases, liquids, solids). It
is not assumed that the particles are identical. Also, the nature of
the interacting forces can be different for different pairs of
particles. Therefore it is not necessary that the particles are
molecules in the sense of chemistry. The theory is also valid when
the continuum is viewed as the system of its atoms (or even
elementary particles). It appears to us physically justified to
assume that atoms (or at least the elementary particles) are formed
of points and rotational and internal degrees of freedom can be
neglected. Hence, for the general theory there is no difference
between a mixture and a chemical compound. The difference is solely
due to the nature of the mutual potentials. The failure of classical
mechanics in the atomistic realm and the influence of quantum
mechanical effects, however, are not taken into account in our
considerations.

In section~2  we will formulate the problem precisely and summarize the
results. By introducing appropriate abbreviations in section~1 the
considerable difficulties in assignment will be eased. In section~3 the
sufficient conditions will be formulated under which the following
investigations are valid. Sections~4 and 5 contain the derivation of the
expected values of the state functions. In sections~6 and 7 we will
prove the validity of the fundamental equations, first for the case
that no external forces are present. In section~8 we will discuss the
influence of external forces. Our investigations are crucially based
on two mathematical Lemmas, formulated and proved in section~9.

The problem being treated here was tackled by Irving and Kirkwood
\cite{irki:50}. This work differs from \cite{irki:50} in the
following points:
\begin{itemize}
\item[1)] The proofs satisfy all requirements of mathematical rigor and, additionally, should be easier than the ones in \cite{irki:50}.
\item[2)] For the stress tensor and the heat flux we give closed integral expressions.
In \cite{irki:50} these quantities are presented as infinite series
which only makes sense if the probability density as a function of
the spatial variable is analytical everywhere.
\item[3)] The interpretation of stress tensor and heat flux as expected values will be carried out in detail.
This is only attempted in \cite{irki:50}.
\item[4)] The assumption that all particles are identical is not made.
\item[5)] The external forces do not have to have a potential and may---apart from space and time---depend also on the velocities of the particles.
\item[6)] The use of $\delta$ functions has been avoided.
In \cite{irki:50} they only serve technical issues.
\end{itemize}

\section{Definitions and assignments}
{\em a. Vectors and tensors.} Vectors and points (point vectors) are
described by lower case letters in bold font, tensors of higher
order are described by capital letters in bold font. The product of
two tensors (dyad) will be represented by the symbol $\otimes$, the
inner product of two vectors or of a vector and a tensor will be
represented by a dot ``$\cdot$'', and  $\mathbf{\nabla_x F(x)}$ represents the
differential (gradient) of the tensor $\mathbf{F(x)}$.
$\mathbf{\nabla_x \cdot F(x)}$ represents the differential (divergence) of $\mathbf{F(x)}$.\\

{\em b. Probability density.} We consider a system of point
particles. The $j^{th}$ particle is ($j$), its mass is $m_j$. A
state of this system is characterized by the particle locations
$\mathbf{x}_1, \cdots , \mathbf{x}_N$ and velocities $\bm{\xi}_1, \cdots ,
\bm{\xi}_N$, {\em i.e.} indicated by a ``point'' ($\mathbf{x}_1, \cdots
, \mathbf{x}_N; \bm{\xi}_1, \cdots , \bm{\xi}_N$) = ($\mathbf{x}_i;
\bm{\xi}_i$) of $6N$-dimensional phase space $\Omega$. This phase
space is the product of $2N$ three dimensional Euclidean vector
spaces\footnote{The $\mathbf{x}_i$ and $\bm{\xi}_i$ vary in the entire,
infinite three dimensional space and not only in a subregion of
it.}. The probability density of the states in $\Omega$ at time $t$
is denoted by
\[
W = W(\mathbf{x}_1, \cdots , \mathbf{x}_N; \bm{\xi}_1, \cdots ,
\bm{\xi}_N; t) = W(\mathbf{x}_i; \bm{\xi}_i;t).
\]

{\em c. Potential energy.} We assume that the force exerted by ($j$)
on ($k$), $\mathbf{k}_{jk}$, depends only on the locations $\mathbf{x}_j$
and $\mathbf{x}_k$. By invariance it follows that $\mathbf{k}_{jk}$ must
be a central force whose contribution is dependent only on the
distance $r_{jk} = | \mathbf{x}_j - \mathbf{x}_k|$. Newton's third law
says that $\mathbf{k}_{jk} = -\mathbf{k}_{kj}$. The particle pair
($j$),($k$) therefore corresponds to a function,
\begin{equation}
V_{jk}(r) = V_{kj}(r),
\end{equation}
such that $V_{jk}(r_{jk})$ yields the potential energy of the pair.
Furthermore,
\begin{equation} \label{1.2}
\begin{array}{rcl}
\mathbf{k}_{jk}& =& - \nabla_{\mathbf{x}_j} V_{jk}(r_{jk}) = -V_{jk}^\prime(r_{jk}) \cfrac{\mathbf{x}_j - \mathbf{x}_k}{r_{jk}} \\
& = &-\mathbf{k}_{kj} = - \nabla_{\mathbf{x}_k} V_{jk}(r_{jk}) =
V_{jk}^\prime(r_{jk}) \cfrac{\mathbf{x}_k - \mathbf{x}_j}{r_{jk}}.
\end{array}
\end{equation}
For the total {\em inner potential energy U} of the system this
results in
\begin{equation}
U = U(\mathbf{x}_1, \cdots , \mathbf{x}_N) =
\sum\limits_{j<k} V_{jk}(r_{jk}) = \frac{1}{2}\sum\limits_{j\neq
k}V_{jk}(r_{jk}).
\end{equation}
Here, one has to sum of all those pairs ($j,k$) for which $j < k$ or
$j \neq k$ is true. It is,
\begin{equation}
 \nabla_{\mathbf{x}_j}U = \sum_{\stackrel{k=1}{
k\neq j}}^N \nabla_{\mathbf{x}_j}V_{jk}(r_{jk}) =
\sum\limits_{\stackrel{k=1}{k\neq j}}^N V^\prime_{jk}(r_{jk})
\frac{\mathbf{x}_j - \mathbf{x}_k}{r_{jk}}. \label{1.4}
\end{equation}

{\em d. External forces.} Apart from those forces acting on particle
($j$) due to other particles we assume that yet another external
force $\mathbf{k}_j$ is exerted that depends only on location ${\bf
x}_j$ and the velocity $\bm{\xi}_j$ of the particle at time $t$,
\[
\mathbf{k}_j = \mathbf{k}_j(\mathbf{x}_j,\bm{\xi}_j,t).
\]
Restrictively, we require, however, that the functions $\mathbf{k}_j$
obey the equation 
\begin{equation}
\sum\limits_{j=1}^N\frac{1}{m_j}\nabla_{\bm{\xi}_j} \cdot {\bf
k}_j(\mathbf{x}_j,\bm{\xi}_j,t) = 0 \label{1.5} 
\end{equation} 
identically. In
particular, this condition is met if the $\mathbf{k}_j({\bf x}_j,\bm{\xi}_j,t)$ do not depend on $\bm{\xi}_j$.

{\em e. Integrals.} The ($6N -3$)-dimensional subspace of $\Omega$,
when discarding the spatial variable $\mathbf{x}_j$ belonging to ($j$),
shall be denoted $\Omega_j$. Analogously, we denote the
subspace $\Omega_{jk}$ that comes from discarding $\mathbf{x}_j$ and
$\mathbf{x}_k$. Let $F$ be a scalar, vector, or tensor function
defined over $\Omega$. Then, the abbreviations,
\[
\langle F | \mathbf{x}_j = \mathbf{x}\rangle, \quad \langle F | \mathbf{x}_j
= \mathbf{x}, \mathbf{x}_k = \mathbf{y}\rangle,
\]
mean that $F$ is the integral over $\Omega_j$ ($\Omega_{jk}$,
respectively), where, after performing the integration, the free
variable $\mathbf{x}_j$ (or the variables $\mathbf{x}_j$ and $\mathbf{x}_k$,
respectively) are to be replaced by $\mathbf{x}$ (or by $\mathbf{x}$ and
$\mathbf{y}$), respectively. Generally, we call \begin{equation} \int_\mathbf{y}f({\bf
y}) \; d\mathbf{y} \nonumber \end{equation} the volume integral of f($\mathbf{y}$)
over the infinite three-dimensional space of locations $\mathbf{y}$.
Then, apparently, the relation
\begin{equation}\label{1.6}
    \int_\mathbf{y}\langle F | \mathbf{x}_j = \mathbf{x}, {\bf
x}_k = \mathbf{y}\rangle \; d\mathbf{y} = \langle F | \mathbf{x}_j = {\bf
x}\rangle
\end{equation}

is true.

\section[Posing the problem]{Posing the problem}
We assume that the probability density, $W(\fatx_i;\bm{\xi}_i;t)$,
is defined for all ($\fatx_i,\bm{\xi}_i) \in \Omega$ and is
continuously differentiable with respect to all variables. Under the
restriction that the external forces fulfill the
condition (\ref{1.5})\footnote{This requirement ensures that phase space is locally volume 
preserving.} the {\em principle of
the conservation of probability} in phase space yields the
differential equation in conventional fashion,
\begin{equation}
\frac{\partial W}{\partial t}  = \sum\limits_{i=1}^N
\left\{-\bm{\xi}_i\cdot \nabla_{\fatx_i}W + \frac{1}{m_i}
(\nabla_{\fatx_i} U - \fatk_i) \cdot \nabla_{\bm{\xi}_i} W \right\},
\label{2.1}
\end{equation}
which determines the rate of change of $W$.

It is the task of this study to derive the fundamental equations of
thermomechanics from (\ref{2.1}) under the regularity
requirements that will be formulated in section~3. First we will assume
the absence of external forces so that (\ref{2.1}) is
simplified to
\begin{equation}
\frac{\partial W}{\partial t}  = \sum\limits_{i=1}^N
\left\{-\bm{\xi}_i\cdot \nabla_{\fatx_i}W + \frac{1}{m_i}
\nabla_{\fatx_i} U \cdot \nabla_{\bm{\xi}_i} W \right\}
\tag{\ref{2.1}a}.
\end{equation}
The case of $\fatk_i \neq 0$ will be treated in section~8.
For $\fatk_i = 0$ the fundamental equations are as follows:\\
\textbf{A}) {\em Continuity equation}
\begin{equation}
\frac{\partial \rho}{\partial t} + \nabla_\fatx \cdot(\rho \fatu) =
0.\label{2.2}
\end{equation}
 \textbf{B}) {\em Equation of motion}
\begin{equation}
\rho\left(\frac{\partial \fatu}{\partial t} + \fatu \cdot
\nabla_\fatx \fatu \right) = \nabla_\fatx \cdot \mathbf{S}. \label{2.3}
\end{equation}
\textbf{C}) {\em Equation of energy}
\begin{equation}
\frac{\partial \epsilon}{\partial t} + \nabla_\fatx \cdot (\fatq -
\fatS\cdot\fatu + \epsilon\fatu) = 0.\label{2.4}
\end{equation}
These fundamental equations connect the following macroscopic state
functions:
\begin{align*}
\rho & = \rho(\fatx, t) = {\text {\em Mass density}},\\
\fatu & = \fatu(\fatx,t) = {\text {\em Velocity}}, \\
\fatS & = \fatS(\fatx,t) = {\text {\em Stress tensor}}, \\
\epsilon & = \epsilon(\fatx,t) = {\text {\em Energy density}},\\
\mathbf{q} & = \mathbf{q}(\mathbf{x},t) = {\text {\em Heat flux
density}}.
\end{align*}

We will prove that these fundamental equations are true if the
aforementioned state functions are interpreted as expected
values. In sections~4--5 we will show that these expected values
are given by the following expressions,
\begin{align}
\rho = \sum\limits_j
m_j\langle W \; | \; \fatx_j = \fatx \rangle,
\label{2.5}\\
\rho\fatu = \sum\limits_j m_j\langle \bm{\xi}_j W \; | \;\fatx_j =
\fatx \rangle. \label{2.6}
\end{align}
The stress tensor is symmetric and
consists of a {\em kinetic contribution}, $\fatS_{\mathrm{K}}$, and
a {\em interaction contribution}, $\fatS_{\mathrm{V}}$,
\begin{align}
\fatS_{\phantom{K}} &= \fatS_{\mathrm{K}} + \fatS_{\mathrm{V}}, \label{2.7}\\
\fatS_{\mathrm{K}}& = -\sum\limits_j m_j \langle(\bm{\xi}_j-\fatu)
\otimes
(\bm{\xi}_j-\fatu)W \; | \; \fatx_j = \fatx\rangle, \label{2.8}\\
\label{2.9}  \fatS_{\mathrm{V}} & = \\
&\frac{1}{2}\sum\limits_{j\neq k}\int_\fatz \left\{ \frac{\fatz
\otimes \fatz}{|\fatz|}V^\prime_{jk}(|\fatz|) \int_{\alpha=0}^1
\langle W \; | \;  \fatx_j = \fatx + \alpha \fatz, \fatx_k = \fatx -
(1-\alpha)\fatz\rangle d\alpha \right\} d\fatz. \nonumber
\end{align}

The energy density splits into the {\em kinetic energy density},
$\epsilon_{\mathrm{K}}$, and the {\em interaction energy
density}, $\epsilon_{\mathrm{V}}$,
\begin{align}
\epsilon_{\phantom{K}} & =  \epsilon_{\mathrm{K}} + \epsilon_{\mathrm{V}},\label{2.10}\\
\epsilon_{\mathrm{K}} & =  \frac{1}{2}\sum_j m_j \langle \bm{\xi}_j^2 W \; | \;\fatx_j = \fatx \rangle,\label{2.11}\\
\epsilon_{\mathrm{V}} & =  \frac{1}{2}\sum_{j\neq k} \langle
V_{jk}(|\fatx_j - \fatx_k|)W \; | \; \fatx_j = \fatx \rangle.
\label{2.12}
\end{align}
The heat flux has three terms, the {\em kinetic contribution},
$\fatq_{\mathrm{K}}$, the {\em transport contribution},
$\fatq_{\mathrm{T}}$, and the {\em interaction contribution},
$\fatq_{\mathrm{V}}$,
\begin{align}
\fatq_{\phantom{K}} & =  \fatq_{\mathrm{K}} + \fatq_{\mathrm{T}} + \fatq_{\mathrm{V}} \label{2.13}\\
\fatq_{\mathrm{K}} & =  \frac{1}{2} \sum_j m_j \langle|\bm{\xi}_j -
\fatu|^2(\bm{\xi}_j -\fatu)
W \;| \; \fatx_j = \fatx \rangle,\label{2.14}\\
\fatq_{\mathrm{T}} & =
\frac{1}{2} \sum_{j \neq k} \langle (\bm{\xi}_j -\fatu)V_{jk}(|\fatx_j - \fatx_k|)W \; | \; \fatx_j = \fatx \rangle,\label{2.15}\\
\fatq_{\mathrm{V}} & =  - \frac{1}{2} \sum_{j \neq k} \int_\fatz
\left\{ \frac{\fatz}{|\fatz|}\fatz \cdot V_{jk}^\prime(|\fatz|)
\right.
 \label{2.16} \\
&\left. \cdot  \int_{\alpha=0}^1 \left\langle \left(
\frac{\bm{\xi}_j + \bm{\xi}_k}{2}-\fatu \right) W \; | \; \fatx_j =
\fatx + \alpha \fatz, \fatx_k = \fatx -(1-\alpha)\fatz \right\rangle
d\alpha \right\} d\fatz.\nonumber
\end{align}

\section{Regularity conditions}
The expectation expressions occurring in (\ref{2.5})-(\ref{2.16})
are improper integrals over $W$ and $V_{jk}$. It is therefore clear
that certain regularity conditions must be met for $W$ and $V_{jk}$
if the state functions in (\ref{2.5})-(\ref{2.16}) are well-defined
and the functions of $\fatx$ and $t$ are continuously differentiable.
Already the condition $\int_\Omega W d\Omega =$ 1 requires that $W$
approaches zero sufficiently fast as $|\fatx_j| \rightarrow
\infty$ and $|\bm{\xi}_j| \rightarrow \infty$.

The following three conditions are sufficient for the validity of the results in this work:\\
\textbf{A}) There is a number $\delta > 0$ such that the function 
\begin{equation*}
G(\fatx_i;\bm{\xi}_i;t) = W(\fatx_i;\fatx_i;t) \prod\limits_{j=1}^N
|\fatx_j|^{3+\delta} \prod\limits_{k=1}^N |\bm{\xi}_k|^{3+\delta},
\end{equation*}
as well as its derivatives are restricted by a constant solely dependent on $t$.\\
\textbf{B}) The functions $V_{jk}(r)$ are defined for all $r$, continuously differentiable and,
together with their derivatives, are finite.\\
\textbf{C}) The functions $\fatk_j(\fatx, \bm{\xi}, t)$ are defined for
all values of $\fatx, \bm{\xi}, t$, continuously differentiable, and
with the constants $A(t)$ and $B(t)$, solely dependent on time, and satisfy
have,
\begin{equation*}
    |\fatk_j| < A(t) |\bm{\xi}| + B(t),  |\nabla_{\bm{\xi}}
\fatk_j| < A(t)|\bm{\xi}| + B(t).
\end{equation*}

These conditions are
sufficient for the convergence of all the improper integrals,
interchanging the order of integration, and for differentiation and
integration etc.

Furthermore, condition \textbf{A} ensures the validity of the
following {\em Lemma}, which is proved by partial integration:

Suppose that $F(\fatx;\bm{\xi}_i)$ is a continuously differentiable
function defined in $\Omega$ which, with the constants $A$ and $B$,
satisfies the inequalities,
\[
|F| < A\prod\limits_{k=1}^N|\bm{\xi}_k|^3 + B, \;\; |\nabla_{\fatx_j}
F| < A \prod\limits_{k=1}^N |\bm{\xi}_k|^3 + B, \;\;
|\nabla_{\bm{\xi}_j} F| < A \prod\limits_{k=1}^N |\bm{\xi}_k|^3 + B.
\]
Then, these formulas are valid:
\begin{equation}
\int F\nabla_{\fatx_j} W = - \int W\nabla_{\fatx_j} F, \;\;\;\; \int
F\nabla_{\bm{\xi}_j} W = - \int W\nabla_{\bm{\xi}_j} F, \label{3.1}
\end{equation}
where one integrates over a subspace of $\Omega$ that contains the
space of $\fatx_j$ or $\bm{\xi}_j$, respectively, as another
subspace (omitting the volume elements). In particular, if $F$ is
independent of $\fatx_j$ or $\bm{\xi}_j$, one has,
\begin{equation}
  \int F\nabla_{\fatx_j}W = 0  \qquad \text{or}  \qquad  \int F\nabla_{\bm{\xi}_j}W = 0,
\quad \text{respectively.}
\label{3.2}
\end{equation}

{\em Note I.} The equation (\ref{2.1}) is a linear partial
differential equation of first order for $W$. According to generally
known rules $W(\fatx_i; \bm{\xi}_i; t)$ is therefore uniquely
determined for all $t$ if the initial density, $W(\fatx_i;
\bm{\xi}_i; 0) = W_0(\fatx_i; \bm{\xi}_i)$ is prescribed. Therefore
it would be desirable to apply regularity conditions only to
$W_0(\fatx_i;\bm{\xi}_i)$ and not to $W(\fatx_i;\bm{\xi}_i;t)$, and
then prove $\mathbf{A}$ as a property of $W$. The author, however, did
not yet succeed in this.

{\em Note II.} If $W$ is not differentiable for all
($\fatx_i;\bm{\xi}_i$) $\in \Omega$ and $t$ the equation
($\ref{2.1}$) generally loses its meaning. The statistical
mechanics, however, is meaningful if $W$ is only integrable over
$\Omega$. It is possible to generalize our investigations to this
case. The condition $\mathbf{A}$ can then be replaced by the
requirement that $W\prod\limits_{k=1}^N|\bm{\xi}_k|^3$ be integrable
for all $t$ over $\Omega$. Then, the expressions
(\ref{2.5})--(\ref{2.16}) remain meaningful. The state functions,
however, being functions of $\fatx$ and $t$ are no longer
continuously differentiable for all values of $\fatx$ and $t$ such
that the fundamental equations in the form (\ref{2.2})-(\ref{2.4})
in general, have no meaning. They must be replaced by the laws of
conservation of mass, momenta, and energy in finite form.

To carry out this sort of generalization demands considerable
formal work. But this can be avoided by understanding all the
present differentiations in the sense of the theory of distributions
by \textsc{L. Schwartz} \cite{schw:50}. Then, (\ref{2.1}) is
completely equivalent with the principle of conservation of
probability. The fundamental equations (\ref{2.2})--(\ref{2.4}) are
generally valid in the sense of the theory of distributions and also
when the probability density $W$ does not even exist. In this latter
case $W$ and the state functions are to be viewed as measures. The
fundamental equations are valid in the conventional sense only for
those values of $\fatx$ and $t$ for which the state functions
represent continuously differentiable functions. 
One can readily derive the transition conditions at the discontinuity
surfaces (collision and acceleration waves) once (\ref{2.2})--(\ref{2.4}) are understood in the sense of distributions.

\section{Expected values of the state functions}

The value of a physical quantity, $F$, for the particle ($j$) shall
be given by a function, $f_j(\fatx_i;\bm{\xi}_i)$, defined in
$\Omega$. According to the rules of probability calculus and
according to section~1$e$ one has
\begin{equation}
\mathcal{E}_j(F)\;d\fatx = \langle
f_j\;W \; | \; \fatx_j = \fatx\rangle d\fatx
\end{equation}
as the expected
value of $F$ for ($j$) under the condition that ($j$) is in the
volume element $d\fatx$ at $\fatx$. Hence, the density
$\mathcal{E}(F)$ of the expected value of $F$ for all particles
at the position $\fatx$ is \begin{equation} \mathcal{E}(F) = \sum\limits_j
\mathcal{E}_j(F) = \sum\limits_j \langle f_j W \; | \;\fatx_j =
\fatx \rangle. \label{4.2} \end{equation} The expected value of $F$ for
($j$) under the condition that ($j$) is in the volume element
$d\fatx$ at $\fatx$ and ($k$) in the volume element $d\faty$ at
$\faty$ is given by
\begin{equation}\label{4.3}
    \mathcal{E}_{jk}(F)\; d\fatx \; d\faty = \langle f_j\;W \; | \; \fatx_j = \fatx,
\fatx_k = \faty \rangle \;d\fatx \;d\faty.
\end{equation}

{\em a. Density and velocity.} It follows immediately from
(\ref{4.2}) that (\ref{2.5}) and (\ref{2.6}) represent the
expected values for mass and momentum densities.

\begin{figure}[hbtp]
\begin{center}
{\includegraphics[angle=0,width=7cm,height=6cm]{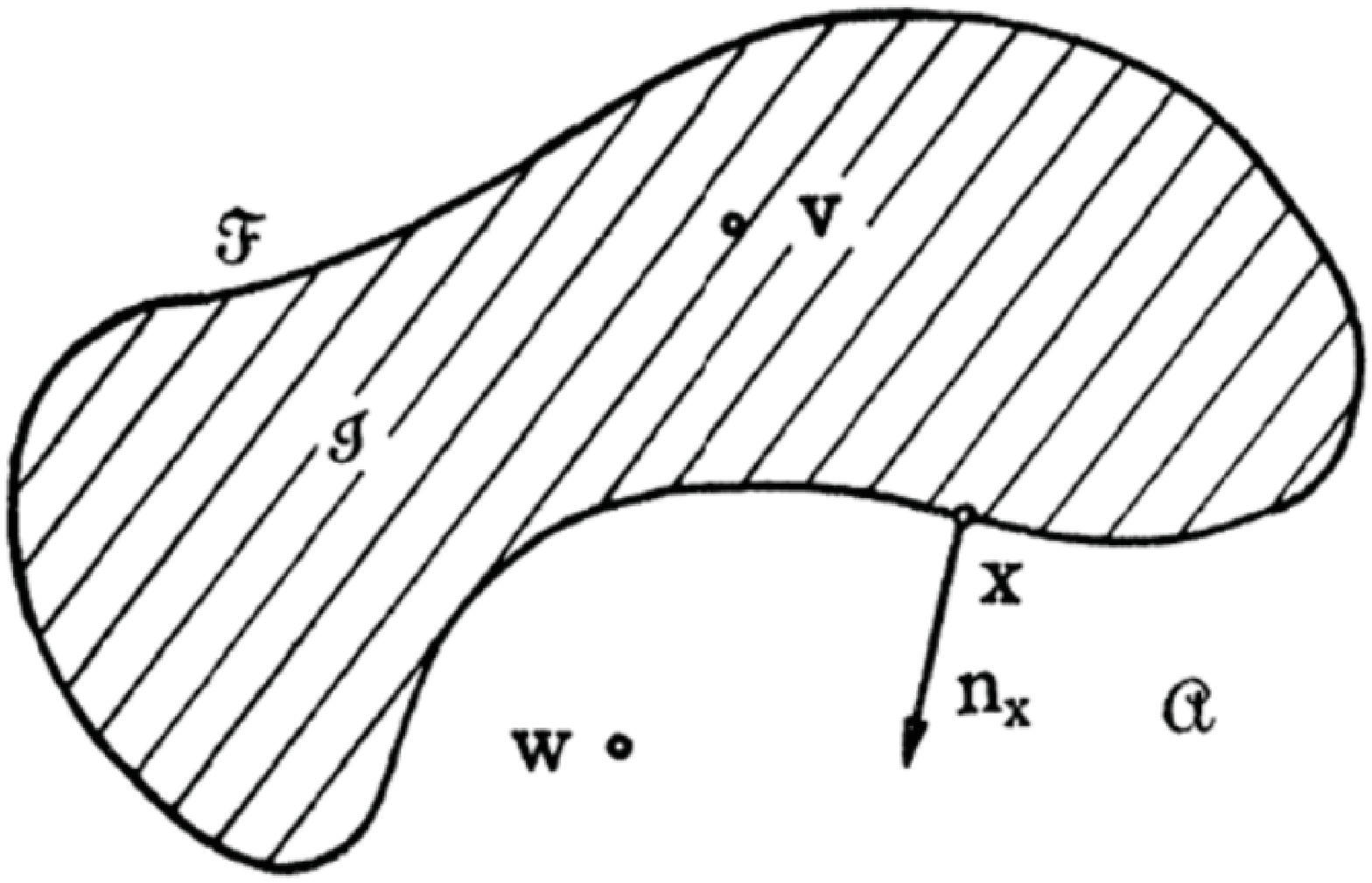}}
\end{center}
\caption{}\label{ABB. 1.}
\end{figure}

{\em b. Stress tensor.} Let $\mathcal{J}$ be a region of
three-dimensional space with a continuously differentiable boundary,
$\mathcal{F}$. We call $\mathcal{A}$ the exterior of
$\mathcal{J}$ with $\fatn_x$ being the normal unit vector at $\fatx
\in \mathcal{F}$ pointing outwards, and with $d\mathcal{F}_x$ the
associated surface element (see Fig. \ref{ABB. 1.}).\footnote{Translator's footnote: The image was cropped from a pdf of the original paper. Note the differences in font styles in the figure and in the text.}

As generally known, the stress tensor is characterized by the fact
that for any part $\mathcal{J}$ of the considered body, the force,
$\fatk$, exerted by $\mathcal{A}$ on $\mathcal{J}$ can be
represented in the form 
\begin{equation}
\fatk = \int_\mathcal{F} \fatS(\fatx)
\cdot \fatn_\fatx \; d\mathcal{F}_\fatx. \label{4.4}
\end{equation}

One readily sees that according to (\ref{2.8}), 
\begin{equation}
\fatk_{\mathrm{K}} = \int_\mathcal{F} \fatS_{\mathrm{K}} \cdot
\fatn_\fatx \; d\mathcal{F}_\fatx \label{4.5} 
\end{equation} 
gives the expected value of the ``kinetic'' force that corresponds to the
momentum per unit time transported from $\mathcal{A}$ to
$\mathcal{J}$. Therefore $\mathcal{J}$ has to be thought of moving
along with the average velocity $\fatu$. The kinetic contribution
(\ref{2.8}) corresponds to what is normally called ``viscous
tension'', in the kinetic theory of mono atomic gases.

Now, let us assume that the particle ($j$) is at the position
$\fatx_j = \fatv \in \mathcal{J}$ and that the particle ($k$) be at
the position $\fatx_k = \fatw \in \mathcal{A}$. Then, according to
(\ref{1.2}), ($k$) exerts the following force on ($j$), 
\begin{equation}
\fatk_{jk}(\fatv,\fatw) = -\nabla_\fatv V_{jk}(|\fatv-\fatw|) = -
V^\prime_{jk}(|\fatv-\fatw|) \frac{\fatv - \fatw}{|\fatv - \fatw|}.
\label{4.6} 
\end{equation} 
According to (\ref{4.3}) the expected value of
the force, exerted by ($k$) in $\mathcal{A}$ on ($j$) in
$\mathcal{J}$, is therefore given by 
\begin{equation*} 
\int_{\fatv \in \mathcal{J}} \int_{\fatw \in \mathcal{A}} \sum\limits_{j \neq k}
\fatk_{jk}(\fatv,\fatw) \langle W \; | \;\fatx_j = \fatv, \fatx_k =
\fatw \rangle \; d\fatv \; d\fatw.
\end{equation*} 
After summation over $j$ and $k$ we obtain from the expected value
$\fatk_{\mathrm{V}}$ of the total force that is exerted by the
particles in $\mathcal{A}$ on those in $\mathcal{J}$. With
(\ref{4.6}) this yields 
\begin{equation*} 
\fatk_{\mathrm{V}} = - \int_{\fatv \in
\mathcal{J}} \int_{\fatw \in \mathcal{A}} \sum\limits_{j \neq k}
\left\{ \frac{\fatv - \fatw}{|\fatv - \fatw|} V^\prime_{jk}(|\fatv -
\fatw|)\langle W \; | \;\fatx_j = \fatv, \fatx_k = \fatw \rangle
\right\} \; d\fatw \; d\fatv.
\end{equation*} 
As one can readily see, the integrand fulfills the conditions $\mathbf{A, B}$ and $\mathbf{C}$ of
section~9. According to Lemma II, (\ref{9.4}) and with (\ref{2.9}) it
therefore follows that
\begin{equation} \fatk_{\mathrm{V}} =
\int_\mathcal{F}\fatS_{\mathrm{V}} \cdot \fatn_\fatx \;
d\mathcal{F}_\fatx. \label{4.7}
\end{equation}
The force $\fatk$ exerted by $\mathcal{A}$ on $\mathcal{J}$ is
composed from the kinetic force $\fatk_{\mathrm{K}}$ (\ref{4.5}) and
the interaction force $\fatk_{\mathrm{V}}$ (\ref{4.7}). Because of
(\ref{2.7}) we therefore have,
\[
\fatk = \fatk_{\mathrm{K}} + \fatk_{\mathrm{V}} = \int_\mathcal{F}
(\fatS_{\mathrm{K}} + \fatS_{\mathrm{V}}) \cdot \fatn_\fatx \;
d\mathcal{F}_\fatx = \int_\mathcal{F} \fatS \cdot \fatn_\fatx \;
d\mathcal{F}_\fatx.
\]
This equation satisfies (\ref{4.4}) and so the expected value
of the stress tensor is indeed given by (\ref{2.7})--(\ref{2.9}).

{\em c. Energy density.} Let us think of the kinetic energy,
$\frac{1}{2}m_j\bm{\xi}_j^2$, of particle ($j$) as being localized
at the position $\fatx_j$ of this particle. Then, according to
(\ref{4.2}) the expected value,
$\epsilon_{\mathrm{K}}(\fatx)$, of the kinetic energy density
at position $\fatx$ is given by (\ref{2.11}).

The localization of the potential energy demands a certain
arbitrariness as it is assigned to particle pairs, and not---as for
the kinetic energy---to individual particles. We assume that the
potential energy corresponding to the pair ($j$), ($k$),
$V_{jk}(|\fatx_j-\fatx_k|)$, is distributed equally at the positions
$\fatx_j$ and $\fatx_k$. Then, at position $\fatx_j$, the total
potential energy is,
\begin{equation}
V_j = V_j(\fatx_1, \cdots, \fatx_N) = V_j(\fatx_i) =
\frac{1}{2}\sum\limits_{\stackrel{k=1}{k\neq j}}^N V_{jk}(r_{jk}),
\;\;\; r_{jk} = |\fatx_j - \fatx_k|. \label{4.8}
\end{equation}
Then, according to (\ref{4.2}), (\ref{2.12}) yields the
expected value, $\epsilon_{\mathrm{V}}(\fatx)$, of the
density localized according to (\ref{4.8}).

{\em d. Heat flux (density).} The heat flux, $\fatq = \fatq(\fatx)$,
is characterized by the fact that for any part $\mathcal{J}$ of the body, the energy transferred per unit time from
$\mathcal{J}$ to $\mathcal{A}$, can be represented in the form, \begin{equation}
Q = \int_\mathcal{F} \fatq(\fatx) \cdot \fatn_\fatx \;
d\mathcal{F}_\fatx, \label{4.9} \end{equation} (see Fig.1).

One is readily convinced that with (\ref{2.14}) the expected
value of the kinetic energy flowing per unit time from
$\mathcal{J}$ to $\mathcal{A}$ is represented by the expression \begin{equation}
Q_{\mathrm{K}} = \int_\mathcal{F} \fatq_{\mathrm{K}} \cdot
\fatn_\fatx \; d\mathcal{F}_\fatx. \label{4.10} \end{equation} Therefore,
$\fatq_{\mathrm{K}}$, given by (\ref{2.14}), is indeed the kinetic
contribution of the heat flux. This is well known from the kinetic
theory of mono-atomic gases.

Let us consider the expression,
\begin{equation}
\fatq^*_{\mathrm{T}} = \fatq_{\mathrm{T}} +
\epsilon_{\mathrm{V}}\fatu = \frac{1}{2} \sum_{j\ne k} \langle
\bm{\xi}_j \; V_{jk}(|\fatx_j - \fatx_k|) W \; | \; \fatx_j =\fatx
\rangle, \label{4.11}
\end{equation}
and compare to (\ref{2.12}) and (\ref{2.15}). According to
(\ref{4.8}) we have \begin{equation} \fatq^*_{\mathrm{T}} = \sum_j\langle
\bm{\xi}_j V_j W \; | \;\fatx_j = \fatx\rangle. \label{4.12} \end{equation}
The expression $ \bm{\xi}_j V_j(\fatx_j)$ indicates the flow of
potential energy, localized according to (\ref{4.8}) at $\fatx_j$.
Therefore, with (\ref{4.12}) and because of (\ref{4.2}), we have
\begin{equation} Q^*_{\mathrm{T}} = \int_\mathcal{F} \fatq^*_{\mathrm{T}} \cdot
\fatn_\fatx \; d\mathcal{F}_\fatx \nonumber \end{equation} as the potential energy that must be transferred per time unit
through $\mathcal{F}$. However, $\mathcal{J}$ has to be seen as
fixed in time. In order to obtain the potential energy
$Q_{\mathrm{T}}$ that is transported per time unit via the moving
surface $\mathcal{F}$ one has to subtract the macroscopic convection
contribution, \begin{equation} Q_{\mathrm{T}}^0 = \int_\mathcal{F}
\epsilon_{\mathrm{V}} \; \fatu \cdot \fatn_\fatx \;
d\mathcal{F}_\fatx. \nonumber \end{equation} So, because of (\ref{4.11}) one
finds \begin{equation} Q_{\mathrm{T}} = Q^*_{\mathrm{T}} - Q^0_{\mathrm{T}} =
\int_\mathcal{F} \fatq_{\mathrm{T}} \cdot \fatn_\fatx
d\mathcal{F}_\fatx. \label{4.13} \end{equation} Therefore, the
$\fatq_{\mathrm{T}}$ given in (\ref{2.15}) is indeed the
contribution of the heat flux that stems from the transport of the
potential energy.

Finally, let us consider
\begin{align}
\fatq_{\mathrm{V}}^* &= \fatq_{\mathrm{V}} - \fatS_{\mathrm{V}} \cdot \fatu \nonumber\\
& = -\frac{1}{2}\sum\limits_{j\ne k} \int_\fatz \left\{ \frac{\fatz}{|\fatz|} \fatz \cdot V_{jk}^\prime(|\fatz|) \right. \label{4.14}\\
& \left. \cdot \int_{\alpha = 0}^1 \left\langle
\frac{\bm{\xi}_j+\bm{\xi}_k}{2} \; W \; | \; \fatx_j = \fatx +
\alpha \fatz, \fatx_k = \fatx -(1-\alpha) \fatz \right\rangle
d\alpha \right\} \; d\fatz, \nonumber
\end{align}
(compare to (\ref{2.9})
and (\ref{2.16})). In section~5 we will show that \begin{equation} Q_{\mathrm{V}}^* =
\int_\mathcal{F} \fatq^*_{\mathrm{V}} \cdot \fatn_\fatx \;
d\mathcal{F}_\fatx \label{4.15} \end{equation} represents the expected
value of the potential energy, transferred per unit time from
$\mathcal{J}$ to $\mathcal{A}$, which stems from particles in
$\mathcal{A}$ performing work on particles in $\mathcal{J}$. Here,
$\mathcal{J}$ can be viewed as fixed in time. In order to obtain the
interaction energy $Q_{\mathrm{V}}$, transferred from $\mathcal{A}$
to the moving $\mathcal{J}$, one has to subtract from
$Q_{\mathrm{V}}^*$ the macroscopic interaction work performed by
$\mathcal{J}$ on $\mathcal{A}$ per unit time, \begin{equation} Q_{\mathrm{V}}^0
= - \int_\mathcal{F} (\fatS_{\mathrm{V}} \cdot \fatu) \cdot
\fatn_\fatx \; d\mathcal{F}_\fatx. \nonumber \end{equation} So, according to
(\ref{4.14}) one finds that \begin{equation} Q_{\mathrm{V}} = Q_{\mathrm{V}}^* -
Q_{\mathrm{V}}^0 = \int_\mathcal{F}(\fatq_{\mathrm{V}}^* +
\fatS_{\mathrm{V}}\cdot \fatu) \cdot \fatn_\fatx \;
d\mathcal{F}_\fatx = \int_\mathcal{F} \fatq_{\mathrm{V}} \cdot
\fatn_\fatx \; d\mathcal{F}_\fatx. \label{4.16} \end{equation} Consequently,
(\ref{2.16}) indeed gives the interaction contribution to the heat
flux.

$Q = Q_{\mathrm{K}}+ Q_{\mathrm{T}} + Q_{\mathrm{V}}$ is the total
energy per unit time transferred from the moving $\mathcal{J}$ to
$\mathcal{A}$, and (\ref{4.9}) is satisfied because of (\ref{4.10}),
(\ref{4.13}), (\ref{4.16}), and (\ref{2.13}). It has therefore been
shown that the expected value of the heat flux is indeed given by
(\ref{2.13})-(\ref{2.16}).

\section{Interaction contribution to the heat flux}

We wish to prove the validity of equation (\ref{4.15}), replacing
expression (\ref{4.14}) by $\fatq_{\mathrm{V}}^*$. To this end we
investigate the mechanical system $\mathcal{S}$ composed solely from
particles in $\mathcal{J}$. For this system, the forces exerted by
the particles in $\mathcal{A}$ on the particles in $\mathcal{J}$ are
considered to be external forces. These forces have the time
dependent variable potential, \begin{equation} \psi(\fatx_i \in \mathcal{J}; t)
= \sum\limits_{\fatx_j \in \mathcal{J}} \sum\limits_{\fatx_k \in
\mathcal{A}} V_{jk}(|\fatx_j - \fatx_k(t)|). \label{5.1} \end{equation} Here,
$\fatx_j \in \mathcal{J}$ is to be viewed as the independent
variable while $\fatx_k(t)$ are those functions that describe the
trajectories of the particles $\fatx_k \in \mathcal{A}$. The internal energy
of the system $\mathcal{S}$ is $E_I$. Then, the total energy $E$
with respect to the external potential (\ref{5.1}) is given by \begin{equation}
E = E_I + \psi(\fatx_i \in \mathcal{J};t). \label{5.2} \end{equation}
According to the energy law in particle mechanics,
\[
\dot{E} = \frac{\partial}{\partial t} \psi(\fatx_i \in \mathcal{J};
t)
\]
along with (\ref{5.1})
\[
\dot{E} = \sum\limits_{\fatx_j \in \mathcal{J}} \sum\limits_{\fatx_k
\in \mathcal{A}} V_{jk}^\prime (|\fatx_j - \fatx_k(t)|) \;
\frac{\fatx_j-\fatx_k(t)}{|\fatx_j - \fatx_k(t)|} \cdot
(-\dot{\fatx}_k(t))
\]
and because of $\dot{\fatx}_k = \bm{\xi}_k$, \begin{equation} \dot{E} =
-\sum\limits_{\fatx_j \in \mathcal{J}} \sum\limits_{\fatx_k \in
\mathcal{A}} V_{jk}^\prime(r_{jk}) \;
\frac{\fatx_j-\fatx_k}{r_{jk}}\cdot \bm{\xi}_k, \;\;\; r_{jk} =
|\fatx_j - \fatx_k|. \label{5.3} \end{equation}

The energy, $E$, however, does not equal that energy that is
localized in $\mathcal{J}$ according to (\ref{4.8}). The
energy $E^*$ localized in $\mathcal{J}$ according to that statement
is composed from the internal energy $E_I$ and half of the potential
energy that corresponds to all those particle pairs (j), (k) for
which $\fatx_j \in \mathcal{J}$ and $\fatx_k \in \mathcal{A}$.
Therefore, we have: \begin{equation} E^* = E_I + \frac{1}{2} \sum_{\fatx_j \in
\mathcal{J}} \sum_{\fatx_k \in \mathcal{A}} V_{jk}(r_{jk}).
\label{5.4} \end{equation} From this, (\ref{5.1}) and (\ref{5.2})
results
\[
E^* = E - \frac{1}{2} \sum_{\fatx_j \in \mathcal{J}} \sum_{\fatx_k
\in \mathcal{A}} V_{jk}(r_{jk}).
\]
Differentiation with respect to $t$ yields \begin{equation} \dot{E}^* = \dot{E}
- \frac{1}{2} \sum_{\fatx_j \in \mathcal{J}} \sum_{\fatx_k \in
\mathcal{A}} V_{jk}^{\prime}(r_{jk}) \;
\frac{\fatx_j-\fatx_k}{r_{jk}}\cdot (\bm{\xi}_j - \bm{\xi}_k),
\label{5.5} \end{equation} where $\dot{\fatx}_j$ and $\dot{\fatx}_k$ have been
replaced by $\bm{\xi}_j$ and $\bm{\xi}_k$, respectively. Insertion
of (\ref{5.3}) into (\ref{5.5}) results in
\[
\dot{E}^* = - \sum_{\fatx_j \in \mathcal{J}} \sum_{\fatx_k \in
\mathcal{A}} V_{jk}^{\prime} (r_{jk}) \;
\frac{\fatx_j-\fatx_k}{r_{jk}}\cdot \frac{\bm{\xi}_j +
\bm{\xi}_k}{2}.
\]
This expression indicates how much the energy, localized by
statement (\ref{4.8}) in $\mathcal{J}$, varies per unit time and due
to the interaction between the particles. The energy flux
contribution directed from $\mathcal{J}$ to $\mathcal{A}$ is
therefore given by $-\dot{E}^*$. According to (\ref{4.3}), the
expected value of this contribution has the form,
\begin{multline*}
Q_{\mathrm{V}}^* = \int_{\fatv \in \mathcal{J}} \int_{\fatw \in
\mathcal{A}} \sum_{j \ne k} V_{jk}^\prime (|\fatv - \fatw|) \;
\frac{\fatv - \fatw}{|\fatv - \fatw|} \\
\cdot \left\langle \frac{\bm{\xi}_j + \bm{\xi}_k}{2} \; W \; | \;
\fatx_j = \fatv, \fatx_k = \fatw \right\rangle \; d\fatv d\fatw.
\end{multline*}

The integrand fulfills the conditions $\mathbf{A, B, C}$ in section~9. With
(\ref{4.14}) this results according to Lemma II, (\ref{9.4}) in
\[
Q_{\mathrm{V}}^* = \int_\mathcal{F} \fatq_{\mathrm{V}}^* \cdot
\fatn_\fatx \; d\mathcal{F}_\fatx.
\]
Q.E.D.

\section{Continuity equation and equation of motion}
The proof of the continuity equation (\ref{2.2}) is very easy.
Multiply (2.1a) with $m_j$, integrate over $\Omega_j$, replace
$\fatx_j$ by $\fatx$ and sum over $j$. Because of (\ref{3.2})
various terms cancel. Eventually, one obtains equation (\ref{2.2})
with (\ref{2.5}) and (\ref{2.6}).

We multiply (2.1a) with $m_j\bm{\xi}_j$, integrate over $\Omega_j$,
replace the free variable $\fatx_j$ by $\fatx$ and finally sum over
$j$, then, taking (\ref{3.2}) and (\ref{2.6}) into account, we
obtain the equation
\begin{align}
& \frac{\partial}{\partial t} (\rho \fatu)  = \fatv_1 + \fatv_2,
\label{6.1} \intertext{where} \fatv_1 &= - \sum_j m_j \langle
\bm{\xi}_j(\bm{\xi}_j \cdot \nabla_{\fatx_j} W) \; | \; \fatx_j =
\fatx \rangle,
\label{6.2}\\
\fatv_2 &= \sum_j \langle (\nabla_{\fatx_j} U \cdot
\nabla_{\bm{\xi}_j} W) \bm{\xi}_j \; | \; \fatx_j = \fatx \rangle.
\label{6.3}
\end{align}
Exchanging integration and differentiation with respect to $\fatx_j$
in (\ref{6.2}) yields, \begin{equation} \fatv_1 = \nabla_\fatx \cdot
\left\{-\sum_j m_j \langle(\bm{\xi}_j \otimes \bm{\xi}_j) W \; | \;
\fatx_j = \fatx \rangle \right\}. \nonumber \end{equation} Since, \begin{equation}
(\bm{\xi}_j \otimes \bm{\xi}_j) = (\bm{\xi}_j - \fatu) \otimes
(\bm{\xi}_j - \fatu ) + \fatu \otimes \bm{\xi}_j + \bm{\xi}_j
\otimes \fatu - \fatu \otimes \fatu, \nonumber \end{equation} and taking
(\ref{2.5}), (\ref{2.6}), and (\ref{2.8}) into account, it follows
that \begin{equation} \fatv_1 = \nabla_\fatx \cdot \{\fatS_{\mathrm{K}} - \fatu
\otimes \rho \fatu \} = \nabla_\fatx \cdot \fatS_{\mathrm{K}} -
\fatu \nabla_\fatx \cdot (\rho \fatu) - \rho \fatu \cdot
\nabla_\fatx \fatu. \label{6.4} \end{equation}

Inserting (\ref{3.1}) in (\ref{6.3}) yields
\[
\fatv_2 = - \sum_j
\langle \nabla_{\fatx_j} U W \; | \;\fatx_j = \fatx \rangle.
\]
If we replace $\nabla_{\fatx_j}U$ by expression
(\ref{1.4}) we find
\[
\fatv_2 = - \sum_{j \ne k} \left\langle V^\prime_{jk}(r_{jk}) \;
\frac{\fatx_j - \fatx_k}{r_{jk}} \; W \; | \;\fatx_j = \fatx
\right\rangle, \;\;\; r_{jk} = |\fatx_j - \fatx_k |.
\]
With (\ref{1.6}),
\begin{equation}\label{6.5}
    \fatv_2 = - \int_\faty \left\{ \frac{\fatx -
\faty}{r} \sum_{j \ne k}  V^\prime_{jk}(r) \langle W \; | \;\fatx_j
= \fatx, \fatx_k = \faty \rangle \right\} d\faty, \;\;\; r = |\fatx
- \faty|.
\end{equation}

One can easily see that the integrand fulfills the conditions ${\bf
A, B, C}$ in section~9. In particular, the validity of (\ref{9.1}) is
found if the summation indices $j$ and $k$ are exchanged in
(\ref{6.5}). According to Lemma I, (\ref{9.2}) follows with
(\ref{2.9}),
\[
\fatv_2 = \nabla_\fatx \cdot \fatS_{\mathrm{V}}.
\]
We insert this and (\ref{6.4}) into (\ref{6.1}), and obtain,
\[
\frac{\partial}{\partial t} (\rho \fatu) = \fatu \frac{\partial
\rho}{\partial t} + \rho \frac{\partial \fatu}{\partial t} = - \fatu
\nabla_\fatx \cdot (\rho \fatu) - \rho \fatu \cdot \nabla_\fatx
\fatu + \nabla_\fatx \cdot (\fatS_{\mathrm{K}} +
\fatS_{\mathrm{V}}).
\]
Bearing in mind the continuity equation (\ref{2.2}) the equation of
motion (\ref{2.3}) indeed follows.

\section{Energy equation}
We multiply (2.1a) with ($1/2 m_j \bm{\xi}_j^2 + V_j)$ (see
(\ref{4.8}), integrate over $\Omega_j$, subsequently replace
$\fatx_j$ by $\fatx$ and sum over $j$. Because of (\ref{3.2})
various terms cancel. Eventually, with (\ref{2.10})-(\ref{2.12}) we
obtain
\begin{align}
&\frac{\partial \epsilon}{\partial t}  = q_1 + q_2 + q_3,
\label{7.1} \intertext{where}
q_1 & = - \frac{1}{2} \sum_j m_j \langle \bm{\xi}_j^2 \bm{\xi}_j \cdot \nabla_{\fatx_j} W \; | \;\fatx_j = \fatx \rangle, \label{7.2}\\
q_2 & = - \sum_j \sum_l \langle V_j \bm{\xi}_l \cdot \nabla_{\fatx_j} W \; | \;\fatx_j = \fatx \rangle, \label{7.3}\\
q_3 & = - \frac{1}{2} \sum_j \langle \bm{\xi}_j^2 \nabla_{\fatx_j} U
\cdot \nabla_{\bm{\xi}_j} W \; | \;\fatx_j = \fatx \rangle.
\label{7.4}
\end{align}

Exchanging integration and differentiation with respect to $\fatx_j$
in (\ref{7.2}) yields
\[
q_1 = -\nabla_\fatx \cdot \{\frac{1}{2} \sum_j m_j \langle
\bm{\xi}_j^2 \bm{\xi}_j W \; | \;\fatx_j = \fatx \rangle \}.
\]
Because
\[
\bm{\xi}_j^2 \bm{\xi}_j = (\bm{\xi}_j - \fatu)^2(\bm{\xi}_j - \fatu)
+ 2 \left[(\bm{\xi}_j - \fatu) \otimes (\bm{\xi}_j - \fatu) \right]
\cdot \fatu + \fatu \bm{\xi}_j^2 + \fatu^2(\bm{\xi}_j - \fatu),
\]
and taking (\ref{2.5}), (\ref{2.6}), (\ref{2.8}), (\ref{2.11}) and
(\ref{2.14}) into account, it follows that \begin{equation} q_1 = - \nabla_\fatx
\cdot(q_{\mathrm{K}} - \fatu \cdot \fatS_{\mathrm{K}} + \fatu
\epsilon_{\mathrm{K}}). \label{7.5} \end{equation}

Inserting (\ref{4.8}) into (\ref{7.3}) and bearing (\ref{3.2}) in
mind, we obtain
\begin{multline} \label{7.6}
q_2 = -\frac{1}{2} \sum_{j \ne k} 
\{ \langle V_{jk} (r_{jk}) \bm{\xi}_j \cdot \nabla_{\fatx_j} W
\; | \; \fatx_j = \fatx \rangle \\
+ \langle V_{jk} (r_{jk}) \bm{\xi}_k \cdot \nabla_{\fatx_k} W \; |
\;\fatx_j = \fatx \rangle \}
\end{multline}
According to the product rule, \begin{equation} V_{jk}(r_{jk}) \nabla_{\fatx_j}
W = \nabla_{\fatx_j} \left[ V_{jk}(r_{jk})W\right] -
\nabla_{\fatx_j} V_{jk}(r_{jk}) W. \nonumber \end{equation} If one inserts
this into the first term on the right hand side of (\ref{7.6}) and
rearranges the second term according to equation (\ref{3.1}) this
yields
\begin{multline} \label{7.7}
q_2 = -\frac{1}{2} \sum_{j \ne k} \{ -\nabla_\fatx \langle \bm{\xi}_j V_{jk} (r_{jk}) W \; | \;\fatx_j = \fatx \rangle \\
+ \langle \bm{\xi}_j \cdot \nabla_{\fatx_j} V_{jk} (r_{jk}) W \; |
\;\fatx_j = \fatx \rangle
 + \langle \bm{\xi}_k \cdot \nabla_{\fatx_k} V_{jk} (r_{jk}) W \; | \;\fatx_j = \fatx \rangle
 \}.
\end{multline}

Equation (\ref{3.1}), applied to (\ref{7.4}), yields
\begin{equation}
    q_3 = -
\sum_j \langle \bm{\xi}_j \cdot \nabla_{\fatx_j} U W \; | \;\fatx_j
= \fatx \rangle = - \sum_{j \ne k} \langle \bm{\xi}_j \cdot
\nabla_{\fatx_j} V_{jk} (r_{jk}) W \; | \;\fatx_j = \fatx \rangle,
\label{7.8}
\end{equation}

 where
(\ref{1.4}) has been employed. Via addition of (\ref{7.7}) and
(\ref{7.8}) and by insertion of (\ref{1.2}) we obtain \begin{equation} q_2 + q_3
= -\nabla_\fatx \cdot \fatq_{\mathrm{T}}^* - q_0, \label{7.9} \end{equation}
where $\fatq_{\mathrm{T}}^*$ is given by (\ref{4.11}) and where \begin{equation}
q_0 = \frac{1}{2} \sum_{j \ne k} \left\langle V_{jk}^\prime(r_{jk})
\frac{\fatx_j - \fatx_k}{r_{jk}} \cdot (\bm{\xi}_j + \bm{\xi}_k) W
\; | \;\fatx_j = \fatx \right\rangle. \nonumber \end{equation} Because of
(\ref{1.6}), we can write $q_0$ in the form of
\[
q_0 = \int_\faty\left\{\frac{\fatx-\faty}{r} \cdot \sum_{j \ne k}
V_{jk}^\prime(r) \cdot \left\langle \frac{\bm{\xi}_j +
\bm{\xi}_k}{2} W \; | \;\fatx_j = \fatx, \fatx_k = \faty
\right\rangle \right\} d\faty, \;\;\; r = | \fatx - \faty |.
\]
In the same way as for (\ref{6.5}) one easily sees also here  that
the conditions \textbf{A, B, C} in section~9 are fulfilled for this
integrand. Therefore it follows from Lemma I, (\ref{9.2}) that \begin{equation}
q_0 = -\nabla_\fatx \cdot \fatq_{\mathrm{V}}^* \nonumber\end{equation} is true
where $\fatq_{\mathrm{V}}^*$ is given by (\ref{4.14}). Inserting
this into (\ref{7.9}) it follows with (\ref{7.5}), (\ref{7.1}),
(\ref{4.11}), and (\ref{4.14}) that
\begin{align*}
\frac{\partial \epsilon}{\partial t}  & =  -\nabla_\fatx \cdot \{\fatq_{\mathrm{K}} - \fatu \cdot \fatS_{\mathrm{K}} + \fatu \epsilon_{\mathrm{K}} + \fatq_{\mathrm{T}}^* + \fatq_{\mathrm{V}}^*\} \\
 & = -\nabla_\fatx \cdot \{\fatq_{\mathrm{K}} + \fatq_{\mathrm{T}} + \fatq_{\mathrm{V}} - \fatu \cdot (\fatS_{\mathrm{K}} + \fatS_{\mathrm{V}}) + \fatu (\epsilon_{\mathrm{K}} + \epsilon_{\mathrm{V}})\},
\end{align*}
{\em i.e.} the energy equation (\ref{2.4}).

\section{External forces}
When the external forces $\fatk_j$ are different from zero and
provided that (\ref{1.5}) is fulfilled, the following situation
emerges:

\textbf{a.} The {\em Continuity Equation} (\ref{2.2}) remains valid.

\textbf{b.} According to \begin{equation} \fatf = \fatf(\fatx, t) = \sum_j \langle
\fatk_j W \; | \;\fatx_j = \fatx \rangle \label{8.1}\end{equation} the {\em
Equation of Motion} (\ref{2.3}) transforms into \begin{equation} \rho
\left(\frac{\partial \fatu}{\partial t}  + \fatu \cdot \nabla_\fatx
\fatu\right) = \nabla_\fatx \cdot \fatS + \fatf \label{8.2}\end{equation}
Here, $\fatf(\fatx,t)$ represents the expected value of the external force density, according to (\ref{4.2}).

\textbf{c.} With \begin{equation} A = A(\fatx, t) = \sum_j \langle\bm{\xi}_j \cdot
\fatk _j W | \fatx_j = \fatx \rangle \label{8.3} \end{equation} the {\em
Energy Equation} (\ref{2.4}) becomes
\begin{equation}
\frac{\partial \epsilon }{\partial t} + \nabla_\fatx \cdot
(\fatq - \fatS \cdot \fatu + \epsilon \fatu) = A. \label{8.4}
\end{equation}
Here, and according to (\ref{4.2}),
$A(\fatx, t)$ corresponds to the expected value of the work
performed by the external forces per unit volume and unit time.

To prove \textbf{a} we bear in mind that the equation
(\ref{2.1}) is obtained by adding the term, \begin{equation} - \sum_{l=1}^N
\frac{1}{m_l} \fatk_l \cdot \nabla_{\bm{\xi}_l} W, \nonumber\end{equation} to
the right hand side of (2.1a). According to the explanations in section~6
one has to add to the right hand side of (2.2) the term \begin{equation} -
\sum_j\sum_l \frac{m_j}{m_l} \langle \fatk_l \cdot
\nabla_{\bm{\xi}_l} W \; | \;\fatx_j = \fatx \rangle. \nonumber\end{equation}
Rearranging with the help of (\ref{3.1}), we obtain
\[
 \sum_j m_j
\left\langle \left(\sum_l \frac{1}{m_l} \nabla_{\bm{\xi}_l} \cdot
\fatk_l\right) W \; | \;\fatx_j = \fatx \right\rangle=0.
\]
This, however, entirely disappears according to (\ref{1.5}), whereby
we have proved claim \textbf{a}. Completely analogous calculations
prove claims \textbf{b} and \textbf{c}.

From (\ref{8.1}) and (\ref{8.3}) one can see that the functional
form $\fatf(\fatx,t)$ and $A(\fatx, t)$ do not only depend on the
the external forces $\fatk_j$ but also on $W$, {\em i.e.} the
respective microscopic state of the system. In certain cases
knowledge about certain macroscopic averages over $W$ suffice for
the determination of $\fatf$ and $A$. We will treat two of such
cases: \textbf{A}) {\em Electrical and gravitational fields.} If the
external forces stem from an electric field of strength
$\fate(\fatx,t)$, we have \begin{equation} \fatk_j(\fatx,\xi,t) = e_j
\fate(\fatx,t), \label{8.5} \end{equation} where $e_j$ is the charge of
particle ($j$). For the expected value of the charge density
$\lambda$ and the electric current density $\fati$, we obtain
according to (\ref{4.2})
\begin{align}
\lambda = \lambda (\fatx, t) = \sum_j e_j \langle W \; | \;\fatx_j = \fatx \rangle, \label{8.6}\\
\fati = \fati(\fatx, t) = \sum_j e_j \langle \bm{\xi}_j W \; |
\;\fatx_j = \fatx \rangle.\label{8.7}
\end{align}
With
(\ref{8.5})--(\ref{8.7}), (\ref{8.1}) and (\ref{8.3}) yield:
\begin{align}
\fatf = \lambda \fate, \qquad A = \fati \cdot \fate. \label{8.8}
\end{align}
For the case of a gravitational field, $\fatg(\fatx, t)$, we have to
set $e_j = m_j$ and obtain, because of (\ref{2.5}) and (\ref{2.6}),
\begin{align}
 \fatf = \rho \fate,  \qquad A = \rho \fatu \cdot \fatg. \label{8.9}
\end{align}

\textbf{B}) {\em Magnetic field.} In this case \begin{equation} \fatk_j(\fatx,
\bm{\xi},t) = e_j \bm{\xi} \times \fatb(\fatx, t), \label{8.10} \end{equation}
where $\fatb$ is the magnetic field strength. Because of
\[
\nabla_{\bm{\xi}} \cdot \fatk_j(\fatx,\bm{\xi},t) =
e_j(\text{rot}_{\bm{\xi}} \bm{\xi}) \cdot \fatb = 0
\]
condition (\ref{1.5}) is fulfilled. For a restricted/limited
$\fatb(\fatx, t)$, the $\fatk_j$ satisfies the regularity condition
\textbf{C} in section~3. Inserting (\ref{8.10}) in (\ref{8.1}) and
(\ref{8.3}), and with (\ref{8.7}) borne in mind, we obtain
\begin{align}
\fatf & = \fatb \times \fati,  \qquad A = 0. \label{8.11}
\end{align}

\section{Two Lemmas}
Let $f(\fatv, \fatw)$ be a scalar vector- or tensor function of the
two vector variables $\fatv$ and $\fatw$ that satisfy the following
conditions:

\textbf{A}) $f(\fatv, \fatw)$ is defined for all $\fatv$ and $\fatw$
and continuously differentiable.

\textbf{B}) There exists a $\delta > 0$ such that the function, \begin{equation}
g(\fatv, \fatw) = f(\fatv, \fatw) |\fatv|^{3+\delta}
|\fatw|^{3+\delta}, \nonumber \end{equation} as well as its derivatives are
bounded.

\textbf{C}) The functional relation \begin{equation} f(\fatv, \fatw) = - f(\fatw,
\fatv) \label{9.1} \end{equation} is true.

Under these circumstances the following two Lemmas are valid.

\noindent \textbf{{\em Lemma I.}} \begin{equation} \int_\faty f(\fatx, \faty)
d\faty = - \frac{1}{2} \nabla_\fatx \cdot \int_\fatz \{\fatz \otimes
\int_{\alpha = 0}^1 f(\fatx + \alpha \fatz, \fatx - (1 - \alpha)
\fatz) d\alpha \} d\fatz \label{9.2} \end{equation} {\em Proof:} The
conditions \textbf{A} and \textbf{B} ensure that the absolute convergence of
the improper integrals, occurring in the following, as well as the
validity of the anticipated exchanges of integration sequences etc.
According to (\ref{9.1}) we have, \begin{equation} \int_\faty f(\fatx, \faty)
d\faty = - \int_\faty f(\faty, \fatx) d\faty. \nonumber \end{equation} If we
introduce the new integration variable $\fatz = \fatx - \faty$ into
the left, and $\fatz = \faty - \fatx$ into the right hand integral,
we find
\begin{align}
\int_\faty f(\fatx, \faty) d\faty &= \int_\fatz f(\fatx, \fatx - \fatz) d\fatz \longeq  - \int_\fatz f(\fatx + \fatz, \fatx) d\fatz \nonumber\\
&=  \frac{1}{2} \int_\fatz [f(\fatx, \fatx- \fatz) - f(\fatx +
\fatz, \fatx)] d\fatz. \label{9.3}
\end{align}
According to the chain rule
this gives
\begin{equation*} \nabla_\fatx f(\fatx + \alpha \fatz, \fatx - (1 -
\alpha) \fatz) = \nabla_\fatv f + \nabla_\fatw f
\end{equation*} and
\[
\frac{d}{d\alpha} f(\fatx + \alpha \fatz, \fatx - (1 - \alpha)
\fatz) = \fatz \cdot(\nabla_\fatv f + \nabla_\fatw f)
\]
where we
have to insert on the right hand sides $\fatv = \fatx + \alpha
\fatz$ and $\fatw = \fatx - (1 - \alpha) \fatz$ as arguments of
$\nabla_\fatv f$ and $\nabla_\fatw f$. Therefore, we have
\[
\fatz \cdot \nabla_\fatx f(\fatx + \alpha \fatz, \fatx - (1 - \alpha
) \fatz ) = \frac{d}{d\alpha} f(\fatx + \alpha \fatz, \fatx - (1 -
\alpha) \fatz).
\]
Integration of this equation with respect to $\alpha$ from $\alpha =
0$ to $\alpha = 1$ yields,
\[
\fatz \cdot \nabla_\fatx \int_{\alpha = 0}^1 f(\fatx + \alpha \fatz,
\fatx - (1 - \alpha) \fatz) d \alpha = f(\fatx + \fatz, \fatx) -
f(\fatx, \fatx - \fatz).
\]
Insertion into (\ref{9.3}) yields (\ref{9.2}). Q. E. D.

\textbf{{\em Lemma II.}} {\em Let $\mathcal{J}$ be any region in space
with piece-wise smooth border surface $\mathcal{F}$. Let
$\mathcal{A}$ be the exterior of $\mathcal{J}$ and $\fatn_\fatx$ the
normal unit vector at point $\fatx$ on $\mathcal{F}$ pointing
outwards (see Fig. 1). Then,}
\begin{multline}\label{9.4}
\int_{\fatv \in \mathcal{J}} \int_{\fatw \in \mathcal{A}} f(\fatv, \fatw) \;d\fatw \;d\fatv \\
= -\frac{1}{2} \int_\mathcal{F} \int_\fatz \int_{\alpha = 0}^1
f(\fatx + \alpha \fatz, \fatx - (1-\alpha)\fatz)(\fatz \cdot
\fatn_\fatx) \;d\alpha \;d\fatz \; d\mathcal{F}_\fatx.
\end{multline}
{\em Proof:} First, one sees immediately that because of the
antisymmetry of $f(\fatv, \fatw)$ (\ref{9.1})
\[
\int_{\fatv \in \mathcal{J}} \int_{\fatw \in \mathcal{J}} f(\fatv,
\fatw) \; d\fatv \; d\fatw = 0.
\]
Therefore, we have 
\begin{equation}
\int_{\fatv \in \mathcal{J}} \int_{\fatw \in
\mathcal{A}} f(\fatv, \fatw) \;d\fatv \;d\fatw = \int_{\fatv \in
\mathcal{J}} \int_{\fatw} f(\fatv, \fatw) \;d\fatw
\;d\fatv. \label{9.5} 
\end{equation} 
Now, according to Lemma I, 
\begin{equation}
\int_\fatw f(\fatv,\fatw) \;d\fatw = \nabla_\fatv \cdot
\fatg(\fatv), \label{9.6}
\end{equation}
where
\begin{equation} 
\fatg(\fatv) = -
\frac{1}{2} \int_\fatz \left\{ \fatz \otimes \int_{\alpha=0}^1
f(\fatv + \alpha \fatz, \fatv - (1 - \alpha) \fatz
\;d\alpha\right\}\;d\fatz. \label{9.7}
\end{equation} 
According to Gauss'
theorem,
\[
\int_{\fatv \in \mathcal{J}} \nabla_\fatv \cdot \fatg(\fatv) \;
d\fatv = \int_\mathcal{F} \fatg(\fatx)\cdot \fatn_\fatx
\;d\mathcal{F}_\fatx.
\]
From this and from (\ref{9.5})--(\ref{9.7}) results in the relation
(\ref{9.4}) claimed. Q. E. D.

\bibliographystyle{plain}
\bibliography{Noll55-sand-report}

\section*{Appendix}
The following changes were made to equations in the original
manuscript.
\begin{enumerate}
  \item Equation (\ref{1.4}): replaced $\nabla\mathbf{x}_j$ with $\nabla_{{\bf
  x}_j}$.
  \item Equation (\ref{2.14}): replaced $(\bm{\xi}_j -
\fatu)((\bm{\xi}_j -\fatu)^2$ with $|\bm{\xi}_j -
\fatu|^2(\bm{\xi}_j -\fatu)$.
\item Section 3: replaced
$G(\fatx_i;\bm{\xi}_i;t) = W(\fatx_i;\fatx_i;t) \displaystyle \prod\limits_{j=1}^N
|\fatx_j|^{3+\delta} \prod\limits_{k=1}^N |\bm{\xi}_k|^{6+\delta}$
with\\
$
G(\fatx_i;\bm{\xi}_i;t) = W(\fatx_i;\fatx_i;t) \displaystyle \prod\limits_{j=1}^N
|\fatx_j|^{3+\delta} \prod\limits_{k=1}^N |\bm{\xi}_k|^{3+\delta}
$
\item
Equation \eqref{3.2}$_1$: replaced $F\nabla_{\fatx_j}W = 0$ with $\int F\nabla_{\fatx_j}W = 0$
\item 
Equation (\ref{7.3}): replaced $\nabla_{\fatx_2}$ with $\nabla_{\fatx_j}$.
\item 
Section 8: replaced
$
 \displaystyle \sum_j m_j
\left\langle \left(\sum_l \frac{1}{m_l} \nabla_{\bm{\xi}_l} \cdot
\fatk_l\right) W \; | \;\fatx_j = \fatx \right\rangle
$
with\\
$
\displaystyle \sum_j m_j
\left\langle \left(\sum_l \frac{1}{m_l} \nabla_{\bm{\xi}_l} \cdot
\fatk_l\right) W \; | \;\fatx_j = \fatx \right\rangle=0
$
\item
Proof of Lemma II: replaced $\displaystyle \int_{\fatv \in \mathcal{J}} \int_{\fatw \in \mathcal{A}} f(\fatv, \fatw) \; d\fatv \; d\fatw = 0$ with\\
$\displaystyle \int_{\fatv \in \mathcal{J}} \int_{\fatw \in \mathcal{J}} f(\fatv,
\fatw) \; d\fatv \; d\fatw = 0$
\item
Equation \eqref{9.5}$_2$: replaced
$ \displaystyle  \int_{\fatv \in \mathcal{J}} \int_{\fatw \in \mathcal{A}} f(\fatv, \fatw) \;d\fatw
\;d\fatv$ with \\
$ \displaystyle  \int_{\fatv \in \mathcal{J}} \int_{\fatw} f(\fatv, \fatw) \;d\fatw \;d\fatv$
\end{enumerate}

\end{document}